\documentclass[10pt,fleqn]{article}

\usepackage{epsfig}
\usepackage{subfigure}
\newcommand{\beq}{\begin{equation}}
\newcommand{\eeq}{\end{equation}}
\newcommand{\bea}{\begin{eqnarray}}
\newcommand{\eea}{\end{eqnarray}}
\newcommand{\lfe}{\lefteqn}
\def\lsim{\mathrel{\rlap{\lower4pt\hbox{\hskip1pt$\sim$}}
    \raise1pt\hbox{$<$}}}         
\def\gsim{\mathrel{\rlap{\lower4pt\hbox{\hskip1pt$\sim$}}
    \raise1pt\hbox{$>$}}}         
\def\esim{\mathrel{\rlap{\raise2pt\hbox{$\sim$}}
    \lower1pt\hbox{$-$}}}         

\begin{document}

\begin{titlepage}
\pagestyle{empty}
\title{Neutralino Annihilation\\ into a Photon and a $Z$ Boson}
\author{Piero Ullio\thanks{piero@physto.se} \\
 Department of Theoretical Physics, Uppsala University,\\
 Box 803, SE-751 08 Uppsala, Sweden\\
 and Department of Physics, Stockholm University, \\
 Box 6730,SE-113~85~Stockholm, Sweden
 \and
Lars Bergstr{\"o}m\thanks{lbe@physto.se} \\
 Department of Physics, Stockholm University,\\
 Box 6730, SE-113~85~Stockholm, Sweden }
\date{October 24, 1997}

\maketitle
\smallskip

\begin{abstract}
A full one-loop calculation of neutralino $S$-wave annihilation into
the $Z\gamma$ final state is performed in the minimal supersymmetric
extension of the Standard Model. This process, like the 
similar one with two photons in the final state, may be of importance
for the indirect detection of supersymmetric dark matter through the
very narrow $\gamma$ ray line that would result from neutralino annihilations
in the galactic halo.

We give the complete analytical formulas for this loop-induced process
and treat the case of a pure Higgsino as a first application of our 
expressions. Predictions for the gamma line flux are given for the
halo model which is of the form suggested by Kravtsov et al. and for
the profile proposed by Navarro, Frenk and White.

For heavy neutralinos, the lines  from $2\gamma$ and $Z\gamma$ 
would have indistiguishable energy
in a realistic detector, making the fluxes add and facilitating discovery. 
For lighter
neutralinos, the positions and relative strengths of the two lines
would give valuable information on the nature of the supersymmetric 
dark matter particles.

\end{abstract}
\end{titlepage}

\pagebreak

\section{Introduction}

Although the lightest supersymmetric particle, generally assumed to 
be a neutralino, is an excellent dark matter candidate it is not easy 
to prove the hypothesis that the massive halo of the Milky Way is 
predominantly composed of such particles. Due to the weak 
interaction strength, direct detection of neutralino scattering 
requires detectors of a sensitivity which only recently has 
started to come close to that needed according to theoretical predictions.

In fact, even if there would appear a direct detection signal, it 
would most likely take a long time before it is established as a 
discovery due to the rather featureless recoil spectrum and the weak 
temporal modulations caused by the Earth's motion in the ``wind'' of 
neutralinos.

A similar lack of distinctive signature is plaguing indirect 
detection of antiprotons, positrons and continuum gamma emission from 
neutralino annihilation in the galactic halo. A better signature is 
provided by high-energy neutrinos from neutralino annihilation in
the centre of the Sun or the Earth, but km$^2$-scale neutrino 
detectors may be needed to scan a substantial portion of 
supersymmetric parameter space.\footnote{For a thorough discussion of 
the various detection methods, see \cite{jkg}.}

An excellent signature is given by a gamma ray line from neutralino 
annihilation into a two-body final state with at least one photon. 
Since the neutralinos annihilate with galactic velocities, i.e. 
non-relativistically, the photon in the
process $\tilde{\chi} + \tilde{\chi} \rightarrow \gamma 
+X^0 $ (where $\chi$ is the neutralino and $X^0$ is a neutral particle)
 will be emitted  with energy

\beq
E_{\gamma} = M_{\chi} - \frac{m_X^2}{4\,M_{\chi}}.
\eeq

In a recent paper \cite{lp}, we have performed a full one-loop 
calculation for the case $X^0=\gamma$, i.e.
$\tilde{\chi}^{0}_{1} + \tilde{\chi}^{0}_{1} \rightarrow \gamma 
+\gamma $, where $\tilde{\chi}^{0}_{1}$ is the lightest of the four 
neutralinos in MSSM, the minimal supersymmetric extension of the Standard 
Model of particle physics. There it was shown that the $\gamma\gamma$
process can be at least an order of magnitude larger than earlier simpler 
estimates indicated, especially for the very heavy, pure higgsino case 
(however, even this larger rate is several orders of magnitude below 
that needed to explain the possible structure in the multi-TeV gamma
spectrum recently reported in \cite{strausz}, unless the halo dark matter is 
very clumpy \cite{lp}).

As there soon will exist new gamma-ray detectors, both space-borne 
and on Earth, with an order of magnitude larger detection area than 
that of existing ones, it is of importance to investigate other 
neutralino annihilation channels which give a nearly monoenergetic 
$\gamma$. In the MSSM, the only other such final state is 
$Z^0\gamma$ \cite{BK}. Note that since the annihilation in the halo 
takes place in the $S$ partial wave, helicity conservation forbids the 
$H^0_{i}\gamma$ final state with $i=1,2,3$ representing the three 
neutral Higgs bosons. There may be non-perturbative final states, 
such as $V\gamma$ with $V$ a vector meson made from a $q\bar q$ pair, 
but these  generally have a very small branching ratio \cite{STS,BS}.

As we will see in the case of a pure higgsino, 
the $Z^0\gamma$ and $\gamma\gamma$ cross sections are 
 of about equal magnitude. This gives an interesting way to verify the 
 gamma ray detection of neutralino dark matter: there should be two 
 lines with energy ratio  
\beq
{E_{1\gamma}\over E_{2\gamma}} = 1 - \frac{m_Z^2}{4\,M_{\chi}^2}.
\eeq
For a heavy higgsino, the two lines would hardly be resolvable 
resulting in a further increase of the line strength compared to the 
results of \cite{lp}. For a lower neutralino mass (below a few hundred 
GeV) the two lines would be resolved. The relative strengths of the 
two lines could give a handle on the composition of the 
neutralino. This comes about because 
despite the fact that the two processes are 
closely related there are some differences (e.g. the non-diagonal 
couplings of $Z^0$ to charginos and squarks) which depend on 
composition. We leave the detailed investigation of this for future 
work, however. 

\section{Cross section}

The process of neutralino annihilation into a photon and a Z$^0$ boson
\beq
\tilde{\chi}^{0}_{1} + \tilde{\chi}^{0}_{1} \rightarrow \gamma 
+ Z^0
\eeq 
was briefly discussed in \cite{BK} but is examined here for the
 first time in a full one-loop calculation. We focus on the case of 
massive non-relativistic particles in the initial state, as this is the 
appropriate limit for neutralinos annihilating in the galactic halo (the 
predicted value of the velocity for dark matter particles in the halo is 
of the order of 10$^{-3}$ $c$). The outgoing photons are  nearly 
monochromatic, with  energy 
\beq
E_{\gamma} = M_{\chi} - \frac{m_Z^2}{4\,M_{\chi}}.
\eeq

The steps we follow to compute the cross section are essentially the same 
as those described for the 2 $\gamma$ case. As the P wave is highly 
suppressed, it is a very good approximation to assume that the neutralino 
pair is in a $^{1}$S$_{0}$ state and to calculate the amplitude using the 
corresponding projector~\cite{kuhn}. We notice  that it is convenient 
to choose the non-linear gauge introduced in Ref~\cite{fujikawa} which has 
the peculiarity of having a vanishing coupling for the 
$W^{\pm}\,G^{\mp}\,\gamma$ vertex ($G^{\mp}$ is the unphysical Higgs or 
charged Goldstone boson). We reduce in this way the number of diagrams 
involving $W^+$ and $G^+$ loops. The diagrams giving a non-vanishing 
contribution to the cross section at the one loop level are shown in 
Figs.\,1-4. 
Note that for a generic supersymmetric model the vertices 
$Z^0\,\tilde{f}_i\,\tilde{f}_j$ and $Z^0\,\chi_i^+\,\chi_j^+$ are non 
diagonal respectively in the sfermion and the chargino mass eigenstates.

The computation of the loop diagrams is greatly simplified after realising 
that
the four-point functions which arise from box diagrams can be rewritten as 
linear combinations of three-point functions. This procedure was already 
exploited in~\cite{lp} and can be applied regardless of the final state, 
because the particles in the initial state, being identical particles at 
rest, have equal four-momenta. It is easy to verify that in this case it 
is possible to find a linear combination of three or four factors in the 
denominator of each four-point function which is independent of the 
momentum flowing in the loop and formally reduce the calculation to 
integrals of three-point functions. 

We keep, where possible, the notation introduced in Ref.~\cite{lp}. The 
amplitude of the process is factorized as
\beq
\mathcal{A} =\frac{e}{2 \sqrt{2} \; \pi^2}
 \epsilon\left(\epsilon_{1},\epsilon_{2},k_{1},k_{2} \right)
 \;\tilde{\mathcal{A}}
\eeq
where $\epsilon_{1}$ and $k_{1}$ are the polarization tensor and the 
momentum of the photon in the final state, whereas $\epsilon_{2}$ and 
$k_{2}$ refer to the polarization tensor and the momentum of the $Z^0$. 
The cross section multiplied by $v$, the relative velocity of the 
neutralino pair, is given in terms of $\tilde{\mathcal{A}}$ by the formula 
\beq
 v\sigma_{Z\gamma} = \frac{\alpha}{32\,\pi^4}
\,\frac{\left(M^2_{\chi}-m^2_{Z}/4\right)^3}{M^4_{\chi}}\; \left|\; 
\tilde{\mathcal{A}}\; \right|^{2}\;\;\;.  \label{eq:sigmav}
\eeq 
We have identified four classes of diagrams. The total amplitude is given 
as:  
\begin{eqnarray*}
\tilde{\mathcal{A}}=\tilde{\mathcal{A}}_{f\tilde{f}}+
 \tilde{\mathcal{A}}_{H^+}+\tilde{\mathcal{A}}_{W}+\tilde{\mathcal{A}}_{G}.
\end{eqnarray*}
For each contribution we have separated real and imaginary parts. The 
results are not as compact as those obtained for the process of 
neutralino annihilation into two photons. This is due mainly to the 
presence in each class of diagrams of four mass parameters. The complexity
of the calculation, and the many parameters entering the supersymmetric
models means that it may be difficult to check our results numerically 
in an independent
calculation. Therefore, we want to give all the analytical expressions
despite the rather lengthy formulas. We have, however, chosen a notation
that facilitates a comparison with the previous $2\gamma$ results \cite{lp}
in the limit when $M_Z\to 0$, and only diagonal terms are kept in the
$Z^0\,\tilde{f}_i\,\tilde{f}_j$ and $Z^0\,\chi_i^+\,\chi_j^+$ vertices.

\textbf{1) Contribution of the fermion-sfermion loop diagrams.} 

This class of diagrams is shown in Fig.~1. The sum over $f$ includes the 
quarks and the charged leptons, the sum over 
$\tilde{f}_i$ and $\tilde{f}_j$ the corresponding sfermion mass 
eigenstates.
\bea
\lefteqn{Re\,\tilde{\mathcal{A}}_{f\tilde{f}}=
\sum_{f} \frac{c_f \cdot e_{f}}{M^2_{\chi}-m^2_Z /4} \;\Bigg\{ 
\,\sum_{\tilde{f_i}}\bigg[ 
E_1\, I_{1}^{\;3}\left(a,b,c/4\right) + E_2\, 
\tilde{I}_{2}^{\;3}\left(a,b,c/4\right) }\nonumber \\
&& +\,E_1\,I_{3}^{\;3}\left(a,b,c/4\right) + 
(E_1+E_3+E_4)\,\tilde{I}_{3}^{\;3}\left(a,b,c/4\right) - \frac{S_{f 
\tilde{f}}^{\;b}}{4}\,I_{4}^{\;1}\left(a,b,c\right) \bigg]  \nonumber \\
&&+\,\sum_{\tilde{f_i}\tilde{f_j}}\bigg[\, 
(E_5+E_6+E_7)\,\tilde{I}_{2;1}^{\;4}\left(a,b,d,c/4\right) + 
(E_5+E_6-E_7)\,\tilde{I}_{2;2}^{\;4}\left(a,b,d,c/4\right) \nonumber \\
&&+\,E_8\,I_{3}^{\;3}\left(a,b,c/4\right) + 
E_8\,I_{3}^{\;4}\left(a,b,d,c/4\right) + \frac{S_{f 
\tilde{f}}^{\;a}}{4}\,I_{4}^{\;2}\left(a,b,d,c\right) \bigg] \Bigg\} 
\nonumber 
\\&&+\,\sum_{f} \frac{c_f \cdot e_{f}}{M^2_{\chi}-m^2_Z 
/4}\,E_9\,I_{1}^{\;3}\left(a,b,c/4\right) \label{reff}
\eea

\bea
\lefteqn{Im\,\tilde{\mathcal{A}}_{f\tilde{f}}=\pi\;\frac{c_f \cdot 
e_{f}}{M^2_{\chi}-m^2_Z /4} \;\Bigg\{
- \sum_{f} \Bigg( \,\sum_{\tilde{f_i}} E_1\,+ E_9 \bigg) 
\,J_{1}\left(a,b\right)\;\; \theta \left(1-m^2_{f}\,/\,M^2_{\chi} \right)} 
\nonumber\\
&&+\;\sum_{f}  \bigg[ \bigg( \sum_{\tilde{f_i}} E_1 + E_9 
\bigg)\,J_{2}\left(b,c\right)\,+\,\sum_{\tilde{f_i}} \bigg( 
(E_1+E_3+E_4)\,J_{3}\left(a,b,c\right) \nonumber \\
&&\;\;\;\;\;\;\;\;\;\;\;\;+\,\frac{c}{8}\,\frac{\sqrt{1-4\,b/c}\;S_{f 
\tilde{f}}^{\;b}}{a-c/4}\bigg) \bigg]
\;\; \theta \left(1-4\,m^2_{f}\,/\,m^2_{Z} \right)\Bigg\}
\eea

$e_{f}$ is the charge of the fermion in units of the electron charge 
($-e$), 
$c_f$ is the color factor equal to 3 for quarks and to 1 for leptons. The 
functions $I$ and $\tilde{I}$ which appear in the real part 
arise from the loop integrations and the $J$ functions are 
the corresponding contributions in the imaginary part; their explicit form 
is given in Appendix A. The coefficients E are listed below:
\begin{eqnarray*}
\lfe{E_1\,=\,-\frac{b\;(S_{f \tilde{f}}^{\;b} + S_{f \tilde{f}}^{\;c})
+ 2\,\sqrt{a\,b}\,D_{f \tilde{f}}}{4\,(1+a-b)}} \\ 
\lfe{E_2\,=\,-\,\frac{1}{4}\,\frac{S_{f \tilde{f}}^{\;b}}{1-b+c/4} }\\
\lfe{E_3\,=\,\frac{1}{4}\,\frac{(b-c/4)\;S_{f \tilde{f}}^{\;b}}{1-b+c/4}} 
\\
\lfe{E_4\,=\,-\,\frac{c}{16}\,\frac{(1-b+c/4)\;S_{f \tilde{f}}^{\;b}}
{(a-c/4)^2}} \\
\lfe{E_5\,=\,-\,\frac{1}{8}\,\frac{(1/2+d/2-c/4)\;S_{f \tilde{f}}^{\;a}}
{1/2+d/2-b-c/4} }\\
\lfe{E_6\,=\,-\,\frac{c}{32}\,\frac{(1/2+d/2-b-c/4)\;S_{f \tilde{f}}^{\;a}}
{(a-c/4)^2}} \\
\lfe{E_7\,=\,\frac{1}{16}\,\frac{(1-d)\;S_{f \tilde{f}}^{\;a}}{a-c/4} }\\
\lfe{E_8\,=\,\frac{1}{8}\,\frac{b\;S_{f \tilde{f}}^{\;a}}{1/2+d/2-b-c/4}} 
\\
\lfe{E_9\,=\,\frac{m^2_{f}\;G_{Zf}}{4\,m^2_{Z}} -\frac{m_{f}\,M_{\chi}\;
   G_{H^{0}_{3} f}}{4\,(4\,M^2_{\chi}-m^2_{H^{0}_{3}} )}}
\end{eqnarray*}
where we have defined:
\begin{eqnarray*}
a=\frac{M^2_{\chi^0_1}}{M^2_{\tilde{f}_i}} &&
b=\frac{m^2_{f}}{M^2_{\tilde{f}_i}} \\
c=\frac{m^2_{Z}}{M^2_{\tilde{f}_i}} &&
d=\frac{M^2_{\tilde{f}_j}}{M^2_{\tilde{f}_i}}
\end{eqnarray*}

\begin{eqnarray*}
\lefteqn{D_{f \tilde{f}}=\frac{1}{4}\;\left(g^L_{\tilde{f}_i f1}\; 
g^{R\,\ast}_{\tilde{f}_i f1}+g^R_{\tilde{f}_i f1}\; 
g^{L\,\ast}_{\tilde{f}_i f1} \right)\cdot\left(g^L_{Zff}+g^R_{Zff}\right)} 
\\ 
\lefteqn{S_{f \tilde{f}}^{\;a}=\frac{g_{Z\tilde{f}_i \tilde{f}_j}}{2}\;
\left( g^L_{\tilde{f}_j f1}\;g^{L\,\ast}_{\tilde{f}_i f1}+g^R_{\tilde{f}_j 
f1}
\; g^{R\,\ast}_{\tilde{f}_i f1} \right)} \\
\lefteqn{S_{f \tilde{f}}^{\;b}=\frac{1}{2}\;\left( g^L_{\tilde{f}_i f1}\;
g^{L\,\ast}_{\tilde{f}_i f1}\;g^{R}_{Zff}+g^R_{\tilde{f}_i f1}\; 
g^{R\,\ast}_{\tilde{f}_i f1}\;g^{L}_{Zff} \right)} \\ 
\lefteqn{S_{f \tilde{f}}^{\;c}=\frac{1}{2}\;\left( g^L_{\tilde{f}_i f1}\;
g^{L\,\ast}_{\tilde{f}_i f1}\;g^{L}_{Zff}+g^R_{\tilde{f}_i f1}\; 
g^{R\,\ast}_{\tilde{f}_i f1}\;g^{R}_{Zff} \right)} \\ 
\lefteqn{G_{Z f}=\frac{1}{2}\;\left( g^L_{Z11} - g^R_{Z11}\right)\;
\left({g^{L}_{Zff}}^2 - {g^{R}_{Zff}}^2\right)} \\ 
\lefteqn{G_{H^{0}_{3} f}=\frac{1}{2}\;\left( g^L_{H^{0}_{3}11} 
- g^R_{H^{0}_{3}11}\right)\;\left(g^L_{H^{0}_{3}ff} - 
g^R_{H^{0}_{3}ff}\right)
\;\left(g^L_{Zff} + g^R_{Zff}\right)} 
\end{eqnarray*}
The index 1 is referred to \(\tilde{\chi}^{0}_{1}\).

\textbf{2) Contribution of the chargino-Higgs boson loop diagrams.}

This class of diagrams is shown in Fig.~2. The sums over $\chi^+_i$ and 
$\chi^+_j$ involves the two chargino mass eigenstates.

\bea
\lefteqn{Re\,\tilde{\mathcal{A}}_{H^+}=
\sum_{\chi^+_i} \frac{1}{M^2_{\chi}-m^2_Z /4}\; \bigg\{ \bigg[
F_1\,I_{2}^{\;3}\left(a,b,c/4\right) +(F_2+F_3)\,
\tilde{I}_{3}^{\;3}\left(a,b,c/4\right)} \nonumber \\
&&+\,\frac{S_{\chi H}^{\;a}}{4}\;I_{4}^{\;1}\left(a,b,c\right) \bigg]
+\,\sum_{\chi^+_j} \bigg[\,(F_4+F_{10})\,I_{1}^{\;4}\left(a,d,1,c/4\right) 
  \nonumber \\
&&+ (F_5+F_{11})\,I_{1}^{\;4}\left(a,1,d,c/4\right) +\,F_4\,I_{2}^{\;3}\left(a,b,c/4\right) \nonumber \\
&&+\, F_5\,I_{2}^{\;4}\left(a,b,d,c/4\right) + (F_5+F_6+F_7+F_8)
\,\tilde{I}_{2;1}^{\;4}\left(a,b,d,c/4\right) \nonumber \\
&&+\,(F_4-F_6+F_7+F_8)\,\tilde{I}_{2;2}^{\;4}\left(a,b,d,c/4\right)
+ F_9\,I_{3}^{\;3}\left(a,b,c/4\right) \nonumber \\
&&+\,F_9\,I_{3}^{\;4}\left(a,b,d,c/4\right)\;-\;\frac{S_{\chi H}^{\;b}}{4}
\;I_{4}^{\;2}\left(a,b,d,c\right)\,\bigg] \bigg\} 
\eea

The coefficients F are listed below:
\begin{eqnarray*}
\lfe{F_1\,=\,-\,\frac{1}{4}\,\frac{S_{\chi H}^{\;a}}{1-b+c/4}} \\ 
\lfe{F_2\,=\,\frac{1}{4}\,\frac{(b-c/4)\;S_{\chi H}^{\;a}}{1-b+c/4}} \\ 
\lfe{F_3\,=\,\frac{c}{16}\,\frac{(1-b+c/4)\;S_{\chi H}^{\;a}}
{(a-c/4)^2}} \\ 
\lfe{F_4\,=\,\frac{1}{8}\,\frac{2\,\sqrt{a}\;D_{\chi H}^{\;b} + 
\sqrt{d}\;S_{\chi H}^{\;c} + S_{\chi H}^{\;b}}{1/2+d/2-a-b}} \\
\lfe{F_5\,=\,\frac{1}{8}\,\frac{2\,\sqrt{a}\,\sqrt{d}\;D_{\chi H}^{\;c} + 
\sqrt{d}\;S_{\chi H}^{\;c} + d\;S_{\chi H}^{\;b}}{1/2+d/2-a-b}} \\
\lfe{F_6\,=\,-\,\frac{1}{8}\,\frac{\sqrt{a}\;D_{\chi H}^{\;b} - 
\sqrt{a}\,\sqrt{d}\;D_{\chi H}^{\;c}}{a-c/4} - \frac{1}{16}\,
\frac{(1-d)\;S_{\chi H}^{\;b}}{a-c/4}} \\
\lfe{F_7\,=\,\frac{c}{32}\,\frac{(1/2+d/2-b-c/4)\;S_{\chi H}^{\;b}}
{(a-c/4)^2}} \\
\lfe{F_8\,=\,-\,\frac{1}{8}\,\frac{(1/2+d/2-c/4)\;S_{\chi H}^{\;b}}
{(1/2+d/2-b-c/4)}} \\
\lfe{F_9\,=\,\frac{1}{8}\,\frac{b\;S_{\chi H}^{\;b}}
{(1/2+d/2-b-c/4)}} \\
\lfe{F_{10}\,=\,\frac{1}{16}\,\frac{\sqrt{d}\;D_Z + S_Z}{c}\,+\,
\frac{1}{8}\,\frac{\sqrt{a}\;S_{H^0}}
{4\,a - m^2_{H^{0}_{3}}/M^2_{\chi^+_i} }} \\
\lfe{F_{11}\,=\,\frac{1}{16}\,\frac{\sqrt{d}\;D_Z + d\;S_Z}{c}\,+\,
\frac{1}{8}\,\frac{\sqrt{a}\,\sqrt{d}\;D_{H^0}}
{4\,a - m^2_{H^{0}_{3}}/M^2_{\chi^+_i}}}
\end{eqnarray*}

and we have defined:
\begin{eqnarray*}
a=\frac{M^2_{\chi^0_1}}{M^2_{\chi^+_i}} && 
b=\frac{m^2_{H^+}}{M^2_{\chi^+_i}} \\ 
c=\frac{m^2_{Z}}{M^2_{\chi^+_i}} &&
d=\frac{M^2_{\chi^+_j}}{M^2_{\chi^+_i}} 
\end{eqnarray*}

\begin{eqnarray*}
\lfe{S_{\chi H}^{\;a}=\frac{g_{ZH^+H^+}}{2}\;\left(g^L_{H^+1i}\; 
g^{L\,\ast}_{H^+1i}+g^R_{H^+1i}\; g^{R\,\ast}_{H^+1i} \right)} \\ 
\lfe{S_{\chi H}^{\;b}=\frac{1}{2}\;\left(g^L_{H^+1j}\; 
g^{L\,\ast}_{H^+1i}\;g^L_{Zji}+g^R_{H^+1j}\; g^{R\,\ast}_{H^+1i}
\;g^R_{Zji} \right)} \\ 
\lfe{D_{\chi H}^{\;b}=\frac{1}{2}\;\left(g^L_{H^+1j}\; 
g^{R\,\ast}_{H^+1i}\;g^L_{Zji}+g^R_{H^+1j}\; g^{L\,\ast}_{H^+1i}
\;g^R_{Zji} \right)} \\ 
\lfe{S_{\chi H}^{\;c}=\frac{1}{2}\;\left(g^L_{H^+1j}\; 
g^{L\,\ast}_{H^+1i}\;g^R_{Zji}+g^R_{H^+1j}\; g^{R\,\ast}_{H^+1i}
\;g^L_{Zji} \right)} \\ 
\lfe{D_{\chi H}^{\;c}=\frac{1}{2}\;\left(g^L_{H^+1j}\; 
g^{R\,\ast}_{H^+1i}\;g^R_{Zji}+g^R_{H^+1j}\; g^{L\,\ast}_{H^+1i}
\;g^L_{Zji} \right)} \\
\lfe{S_{Z}=\left(g^L_{Z11}-g^R_{Z11}\right)\;
\left(g^L_{Zij}\;g^L_{Zji}-g^R_{Zij}\;g^R_{Zji}\right)} \\
\lfe{D_{Z}=\left(g^L_{Z11}-g^R_{Z11}\right)\;
\left(g^L_{Zij}\;g^R_{Zji}-g^R_{Zij}\;g^L_{Zji}\right)} \\
\lfe{S_{H^0}=\left(g^R_{H^{0}_{3}11}-g^L_{H^{0}_{3}11}\right)\;
\left(g^L_{H^{0}_{3}ij}\;g^L_{Zji}-g^R_{H^{0}_{3}ij}\;g^R_{Zji}\right)} \\
\lfe{D_{H^0}=\left(g^R_{H^{0}_{3}11}-g^L_{H^{0}_{3}11}\right)\;
\left(g^L_{H^{0}_{3}ij}\;g^R_{Zji}-g^R_{H^{0}_{3}ij}\;g^L_{Zji}\right)} \\
\end{eqnarray*}

where the indices $i$ and $j$ are referred to \(\tilde{\chi}^{+}_{i}\) and
\(\tilde{\chi}^{+}_{i}\)

\textbf{3) Contribution of the chargino-W boson loop diagrams.}

This class of diagrams is shown in Fig.~3.
\bea
\lfe{Re\,\tilde{\mathcal{A}}_{W}= \sum_{\chi^+_i} \frac{1}
{M^2_{\chi}-m^2_Z /4}\; \bigg\{ \bigg[ 
G_1\,I_{1}^{\;3}\left(a,b,c/4\right) 
+ G_2\,I_{2}^{\;3}\left(a,b,c/4\right)} \nonumber \\
&&+\,G_3\,I_{3}^{\;3}\left(a,b,c/4\right) 
+ \left(G_3+G_4+G_9\right)\,\tilde{I}_{3}^{\;3}\left(a,b,c/4\right) 
- \frac{S_{\chi W}^{\;a}}{2}\;I_{4}^{\;1}\left(a,b,c\right) \bigg]
\nonumber \\
&&+\;\sum_{\chi^+_j} \bigg[\,G_5\,I_{1}^{\;4}\left(a,d,1,c/4\right) 
+ G_6\,I_{1}^{\;4}\left(a,1,d,c/4\right) 
+ G_5\,I_{2}^{\;3}\left(a,b,c/4\right) \nonumber \\
&&\;\;\;\;\;\;\;\;\;\;\;\;+\,G_6\,I_{2}^{\;4}\left(a,b,d,c/4\right) +
 \left(G_6+G_7+G_{10}+G_{11}\right)\,\tilde{I}_{2;1}^{\;4}
\left(a,b,d,c/4\right) \nonumber \\
&&\;\;\;\;\;\;\;\;\;\;\;\;+\,\left(G_5+G_7+G_{10}-G_{11}\right)
\,\tilde{I}_{2;2}^{\;4}\left(a,b,d,c/4\right) 
+ G_8\,I_{3}^{\;3}\left(a,b,c/4\right) \nonumber \\
&&\;\;\;\;\;\;\;\;\;\;\;\;+\,G_8\,I_{3}^{\;4}\left(a,b,d,c/4\right) 
- \frac{S_{\chi W}^{\;b}}{2}\;I_{4}^{\;2}\left(a,b,d,c\right)
\,\bigg] \bigg\}
\eea

\bea
\lfe{Im\,\tilde{\mathcal{A}}_{W} = 
 -\pi\;\sum_{\chi^+_i} \frac{1}{M^2_{\chi}-m^2_Z /4}\; 
G_1\,J_{1}\left(a,b\right)\;\; \theta \left(1-m^2_{W}\,/\,M^2_{\chi} 
\right)} 
\eea

The coefficients G are given by:
\begin{eqnarray*}
\lefteqn{G_1\,=\,2\;\frac{\left(a-b\right)\;S_{\chi W}^{\;a}}{1+a-b}} \\ 
\lefteqn{G_2\,=\,-\,\frac{1}{2}\,\frac{3\,S_{\chi W}^{\;a}
-4 \sqrt{a}\;D_{\chi W}^{\;a}}{1-b+c/4}} \\
\lefteqn{G_3\,=\,-\,\frac{\left(1-a+b\right)\;S_{\chi W}^{\;a}}{1+a-b}} \\
\lefteqn{G_4\,=\,\frac{1}{2}\,\frac{\left(2+b-c/4\right)\,S_{\chi W}^{\;a}
-4 \sqrt{a}\;D_{\chi W}^{\;a}}{1-b+c/4}} \\
\lefteqn{G_5\,=\,\frac{1}{2}\,\frac{2\,\sqrt{a}\;D_{\chi W}^{\;b}\,
-\,\left(\sqrt{d}/2+1/2\right)\;S_{\chi W}^{\;b}}{1/2+d/2-a-b}} \\
\lefteqn{G_6\,=\,\frac{1}{2}\,\frac{2\,\sqrt{a}\sqrt{d}\;D_{\chi W}^{\;b}\,
-\,\left(\sqrt{d}/2+d/2\right)\;S_{\chi W}^{\;b}}{1/2+d/2-a-b}} \\
\lefteqn{G_7\,=\,-\,\frac{1}{2}\,\frac{\sqrt{a}\,(\sqrt{d}+1)\;D_{\chi 
W}^{\;b}
\,-\,(\sqrt{d}+d/4+1/4+c/8)\;S_{\chi W}^{\;b}}{1/2+d/2-b-c/4}} \\
\lefteqn{G_8\,=\,\frac{1}{2}\,\frac{\sqrt{a}\,(\sqrt{d}+1)\;D_{\chi 
W}^{\;b}
\,-\,(\sqrt{d}+b/2+c/4)\;S_{\chi W}^{\;b}}{1/2+d/2-b-c/4}} \\
\lefteqn{G_9\,=\,-\,\frac{c}{8}\;\frac{\left(1-b+c/4\right)\;S_{\chi 
W}^{\;a}}
{\left(a-c/4\right)^2}} \\
\lefteqn{G_{10}\,=\,\frac{c}{16}\;\frac{\left(1/2+d/2-b-c/4\right)\;
S_{\chi W}^{\;b}}{\left(a-c/4\right)^2}} \\
\lefteqn{G_{11}\,=\,-\,\frac{1}{8}\;\frac{(1-d)\;S_{\chi W}^{\;b}}
{\left(a-c/4\right)}}
\end{eqnarray*}

where we have defined:
\begin{eqnarray*}
a=\frac{M^2_{\chi^0_1}}{M^2_{\chi^+_i}} && 
b=\frac{m^2_{W}}{M^2_{\chi^+_i}} \\ 
c=\frac{m^2_{Z}}{M^2_{\chi^+_i}} &&
d=\frac{M^2_{\chi^+_j}}{M^2_{\chi^+_i}} 
\end{eqnarray*}

\begin{eqnarray*}
\lefteqn{S_{\chi W}^{\;a}=\frac{g\,\cos{\theta_W}}{2}\;\left(g^L_{W1i}\; 
g^{L\,\ast}_{W1i}+g^R_{W1i}\; g^{R\,\ast}_{W1i} \right)} \\ 
\lefteqn{D_{\chi W}^{\;a}=\frac{g\,\cos{\theta_W}}{2}\;\left( g^L_{W1i}\; 
 g^{R\,\ast}_{W1i}+g^R_{W1i}\; g^{L\,\ast}_{W1i} \right)} \\
\lefteqn{S_{\chi W}^{\;b}=\frac{1}{4}\;\left(g^L_{W1j}\; 
g^{L\,\ast}_{W1i}+g^R_{W1j}\; g^{R\,\ast}_{W1i} \right)\cdot
\left(g^L_{Zji}+g^R_{Zji}\right)} \\ 
\lefteqn{D_{\chi W}^{\;b}=\frac{1}{4}\;\left(g^L_{W1j}\; 
g^{R\,\ast}_{W1i}+g^R_{W1j}\; g^{L\,\ast}_{W1i} \right)\cdot
\left(g^L_{Zji}+g^R_{Zji}\right)}  
\end{eqnarray*}

\textbf{4) Contribution of the chargino-unphysical Higgs loop diagrams.}

This class of diagrams is shown in Fig.~4.
\bea
\lefteqn{Re\,\tilde{\mathcal{A}}_{G}=
\sum_{\chi^+_i} \frac{1}{M^2_{\chi}-m^2_Z /4}\; \bigg\{ \bigg[
\,(H_1+H_{10})\,I_{2}^{\;3}\left(a,b,c/4\right) 
+(H_2+H_3-H_{10})\,\tilde{I}_{3}^{\;3}\left(a,b,c/4\right)} \nonumber \\
&&+\,\frac{S_{\chi G}^{\;a}}{4}\;I_{4}^{\;1}\left(a,b,c\right) \bigg]
+\,\sum_{\chi^+_j} \bigg[\,H_4\,I_{1}^{\;4}\left(a,d,1,c/4\right) 
+ H_5\,I_{1}^{\;4}\left(a,1,d,c/4\right) \nonumber \\
&&+\,H_4\,I_{2}^{\;3}\left(a,b,c/4\right) 
+ H_5\,I_{2}^{\;4}\left(a,b,d,c/4\right) 
+ (H_5+H_6+H_7+H_8)\,\tilde{I}_{2;1}^{\;4}\left(a,b,d,c/4\right) \nonumber 
\\
&&+\,(H_4-H_6+H_7+H_8)\,\tilde{I}_{2;2}^{\;4}\left(a,b,d,c/4\right)
+ H_9\,I_{3}^{\;3}\left(a,b,c/4\right) \nonumber \\
&&+\,H_9\,I_{3}^{\;4}\left(a,b,d,c/4\right)\;
-\;\frac{S_{\chi G}^{\;b}}{4}\;I_{4}^{\;2}\left(a,b,d,c\right)
\,\bigg] \bigg\} 
\eea

For $i\in\{1..9\}$ the coefficient $H_i$ is obtained from the 
corresponding $F_i$ substituting $S_{\chi H}^{\;p}$ and $D_{\chi H}^{\;p}$ 
respectively with $S_{\chi G}^{\;p}$ and $D_{\chi G}^{\;p}$. $H_{10}$ is 
given by:

\begin{eqnarray*}
\lefteqn{H_{10}\,=\,-\,\frac{\sqrt{c}}{4}\;\frac{S_{\chi G}^{\;d} + 
D_{\chi G}^{\;d}}{1-b+c/4}}
\end{eqnarray*}

a,b,c,d are defined as in the case of the chargino-W boson contribution 
and we have introduced:

\begin{eqnarray*}
\lfe{S_{\chi G}^{\;a}=\frac{g_{ZGG}}{2}\;\left(g^L_{G1i}\; 
g^{L\,\ast}_{G1i}+g^R_{G1i}\; g^{R\,\ast}_{G1i} \right)} \\ 
\lfe{S_{\chi G}^{\;b}=\frac{1}{2}\;\left(g^L_{G1j}\; 
g^{L\,\ast}_{G1i}\;g^L_{Zji}+g^R_{G1j}\; g^{R\,\ast}_{G1i}
\;g^R_{Zji} \right)} \\ 
\lfe{D_{\chi G}^{\;b}=\frac{1}{2}\;\left(g^L_{G1j}\; 
g^{R\,\ast}_{G1i}\;g^L_{Zji}+g^R_{G1j}\; g^{L\,\ast}_{G1i}
\;g^R_{Zji} \right)} \\ 
\lfe{S_{\chi G}^{\;c}=\frac{1}{2}\;\left(g^L_{G1j}\; 
g^{L\,\ast}_{G1i}\;g^R_{Zji}+g^R_{G1j}\; g^{R\,\ast}_{G1i}
\;g^L_{Zji} \right)} \\ 
\lfe{D_{\chi G}^{\;c}=\frac{1}{2}\;\left(g^L_{G1j}\; 
g^{R\,\ast}_{G1i}\;g^R_{Zji}+g^R_{G1j}\; g^{L\,\ast}_{G1i}
\;g^L_{Zji} \right)} \\ 
\lfe{S_{\chi G}^{\;d}=\frac{g}{2}\;\left(g^L_{W1i}\; 
g^{L\,\ast}_{G1i}+g^R_{W1i}\; g^{R\,\ast}_{G1i}\right)} \\ 
\lfe{D_{\chi G}^{\;d}=\frac{g}{2}\;\left(g^L_{G1i}\; 
g^{R\,\ast}_{W1i}+g^R_{G1i}\; g^{L\,\ast}_{W1i} \right)}  
\end{eqnarray*}

The coupling constants for left and right chiral states, $g^L$ and $g^R$, 
are written in the conventions adopted in the PhD Thesis of 
Edsj\"{o}~\cite{joakim}; all of them are defined therein but $g^L_{G1i}$ 
and $g^R_{G1i}$ which are given in~\cite{lp} and
\begin{eqnarray*}
g_{ZGG} = g_{ZH^+H^+}
\end{eqnarray*}
Note that a sign factor is implied wherever a square root of a mass 
parameter appears. For instance, in the third class of diagrams
\beq
\sqrt{a} = sign\left(M_{\chi^0_1} / M_{\chi^+_i}\right) \cdot 
\sqrt{M^2_{\chi^0_1} / M^2_{\chi^+_i}}
\eeq
In this way we take into account the fact that, if the convention of real 
neutralino and chargino mixing matrices is adopted, the mass eigenvalues 
can be either positive or negative.

We have checked the result of the imaginary part applying the Cutkosky 
rules. These provide a general scheme to extract the imaginary part which 
is the contribution to the amplitude deriving from physical intermediate 
states in the loop diagrams (for a review see for instance 
Ref~\cite{zuber}). It is sufficient to examine if in each diagram it is 
possible to identify two separate processes that are both kinematically 
allowed.

\section{Pure Higgsino limit}

The calculation of the annihilation rate involves many
subtleties. Although all integral expressions given here are convergent,
they often individually have large values with severe cancellations 
between different terms. We have checked that our results reproduce
the corresponding $2\gamma$ ones in the limit $M_Z\to 0$.

As there has recently been some interest in the TeV region due to
a claim in the literature for a possible structure in the gamma ray
energy spectrum \cite{strausz},  we give as a first application of our
results a treatment of the pure higgsino case\footnote{The demand of not
overclosing the Universe generally means that such a heavy neutralino
has to have a very large higgsino fraction \cite{joakimpaolo}.}, 
leaving a more
complete phenomenological analysis for a future 
work.

We now present the results  for a pure higgsino  initial state.
Apart from small contributions from the fermion diagrams, 
only the diagrams in Fig.~3 give an important
contribution. The mass of the heaviest chargino goes to infinity, so
the indices i and j  refer to the lightest chargino. The latter is 
nearly degenerate in mass with $\chi^0_1$. We find a compact expression 
which can be easily implemented numerically.     
\bea
\lfe{Re\,\tilde{\mathcal{A}}=\frac{1}{M^2_{\chi}-m^2_Z /4}\; 
\bigg[\,C_1\,I_{1}^{\;3}\left(a,b,c/4\right) + 2\,C_5\,I_{1}^{\;3}\left(a,1,c/4\right)} \nonumber \\
&&+\,(C_2+2\,C_5)\,I_{2}^{\;3}\left(a,b,c/4\right)
+\,2\,\left(C_5+C_7+C_{10}\right)\,\tilde{I}_{2}^{\;3}\left(a,b,c/4\right)
\nonumber \\
&&+\,(C_3+2\,C_8)\,I_{3}^{\;3}\left(a,b,c/4\right)
 +\,(C_3+C_4+C_9)\,\tilde{I}_{3}^{\;3}\left(a,b,c/4\right) \nonumber \\
&&- \frac{S_{\chi W}^{\;a}}{2}\;
I_{4}^{\;1}\left(a,b,c\right) - \frac{S_{\chi W}^{\;b}}{2}\;
I_{4}^{\;2}\left(a,b,1,c\right) \bigg]
\eea

\bea
\lfe{Im\,\tilde{\mathcal{A}} = 
 -\pi\, \frac{1}{M^2_{\chi}-m^2_Z /4}\; 
C_1\,J_{1}\left(a,b\right)\;\; \theta \left(1-m^2_{W}\,/\,M^2_{\chi} \right)} 
\eea

The coefficients $C_i$ are obtained from the corresponding $G_i$ with d=1. 
The degeneration in mass between the neutralino and lightest chargino gives $a\simeq 1$, while the couplings are: 
\begin{eqnarray*}
\lefteqn{S_{\chi W}^{\;a}=D_{\chi W}^{\;a}=\frac{g^3\,\cos{\theta_W}}{4}} \\
\lefteqn{S_{\chi W}^{\;b}=D_{\chi W}^{\;b}=\frac{g^3}{4\,\cos{\theta_W}}\,\left(-\,\frac{1}{2}\,+\,\sin^2{\theta_W}\right)} 
\end{eqnarray*}

In Fig.~5, we show the values of $v\sigma_{Z\gamma}$  obtained from these
expressions, as a function of higgsino mass. As can be seen, the maximum
value is around $3.6\cdot 10^{-28}$~cm$^3$s$^{-1}$ for a higgsino mass
around 140 GeV. For higher masses, the value of $v\sigma_{Z\gamma}$
reaches a plateau of around $0.6 \cdot 10^{-28}$~cm$^3$s$^{-1}$. This
interesting effect of a non-diminishing cross section with higgsino mass
(which is due to a contribution to the real part of the amplitude
coming from diagrams 3e and 3f) was found also for the $2\gamma$ final
state in the corresponding limit, which gave a value of
$1\cdot 10^{-28}$~cm$^3$s$^{-1}$ \cite{lp}. In the same Figure, the 
contribution from the imaginary part is also shown. This drops quite fast
for the highest masses, and constitutes a unitarity lower 
bound on the cross section which agrees with the one found in \cite{BK}.

\section{Halo models}
An uncertain element in the calculation of the absolute gamma line
flux from neutralino annihilation in the galactic halo is the detailed
dark matter density distribution. Since the integration along the line
of sight involves the square of the density, any density enhancement
may affect the predicted flux appreciably.

Available observational data give only  very poor constraints on the mass 
distribution within the Milky Way \cite{binney}. Assuming that dark matter 
profiles are of a universal functional form, we will examine two profiles 
derived with N-body simulations of hierarchical clustering in cold dark matter 
cosmologies and fitted to a sample of dark matter dominated dwarf and 
low-surface brightness galaxies. We consider among the general profile family

\beq
\rho(r) \propto \frac{1}{(r/a)^{\gamma}\;[1+(r/a)^{\alpha}]^{(\beta-\gamma)/\alpha}},
\eeq

the Kravtsov et al. profile \cite{kravtsov}, defined by $(\alpha,\beta,\gamma)=(2,3,0.2)$
with a very mild singularity at the galactic centre, and the Navarro et al. 
profile~\cite{navarro}, which has $(\alpha,\beta,\gamma)=(1,3,1)$.
There are models \cite{bere} which have a more singular behaviour
near the galactic centre, and which would give enormously enhanced rate
in that direction. However, there is observational evidence against
such singularities from cluster gravitational lensing and the
rotation curves of dwarf spiral galaxies \cite{flores}.
On the other hand, the discrepancy between the 1/r central cusp of the 
Navarro et al. profile and the experimental data from the dwarf spheroidal 
DDO~154 has been explained in Ref.~\cite{bs} assuming an additional 
component of dark baryons.

We fix the normalization of the two profiles assuming that the dark matter density 
at our galactocentric distance, $R \simeq 8.5\, {\rm kpc}$, is $\rho_0 \simeq 0.3\, {\rm GeV}/{\rm cm}^3$, 
and we choose for the Navarro et al. profile $a \simeq 25\, {\rm kpc}$ (appropriate value 
for the Milky Way in an $\Omega=1$ cosmology \cite{frenk_priv}), while for the 
Kravtsov et al. we fix $a \simeq 11\, {\rm kpc}$.
We consider the profiles valid up to the capture radius of the black hole at the 
galactic centre ( $\sim 0.01\, {\rm pc}$ for a mass of Sgr A$^*$ $ {\rm M} \simeq 10^6\, {\rm M}_{\odot}$).

The gamma ray flux from the $Z\gamma$ process is given by
\bea
\Phi_{\gamma}(\psi) &=& \frac{v\sigma_{Z \gamma}}{4\pi M_\chi^2} 
\int_{line\;of\;sight}\rho^2(l)\; d\,l(\psi) \nonumber \\
&\simeq& 1.87 \cdot 10^{-11}\left( \frac{v\sigma_{Z \gamma}}
{10^{-29}\ {\rm cm}^3 {\rm s}^{-1}}\right) 
\left(\frac{10\, {\rm GeV}}{M_\chi}\right)^2
\left(\frac{R}{8.5\,{\rm kpc}}\right)\nonumber \\
&&\;\; \cdot \left(\frac{\rho_0}{0.3\,{\rm GeV}/{\rm cm}^3}\right)^2 \; J\left(\psi\right)
\,\,\, {\rm cm}^{-2}{\rm s}^{-1}{\rm sr}^{-1}
\eea

where $\psi$ is the angle between the direction of the galactic centre and 
that of observation and $J\left(\psi\right)$ is a dimensionless function defined as

\beq
J\left(\psi\right) = \frac{1} {R\;{\rho_0}^2} 
\int_{line\;of\;sight}\rho^2(l)\; d\,l(\psi).
\eeq

We have computed the function $J(\psi)$ numerically and the result for the two halo 
models considered are given in Fig.~6. As can be seen, for angles larger than 60 degrees 
the two curves nearly coincide and this is true even if we vary the parameters around 
the chosen values.
Towards the galactic centre the Navarro et al. profile gives a huge enhancement of $J$; 
these values should be compared to the value 5.7, which is the maximal value 
(in the $\psi=0$ direction) of the corresponding function for a halo 
described by an isothermal sphere, $(\alpha,\beta,\gamma)=(2,2,0)$, 
with $a = R$ \cite{turner}. For the Kravtsov et al. profile, 
the maximal value is $J(\sim 0) \simeq 10$.

We are now in the position of being able to give a firm prediction for
the total line flux of $\gamma$ rays for TeV-scale higgsinos. Since
any presently conceivable detector will have an energy resolution not
better than one per cent or so, one has to add the contributions from
the $2\gamma$ annihilation calculated in \cite{lp} and the 
$Z\gamma$ process calculated here for the first time. With the Kravtsov et al. density profile, 
the flux in the direction of the galactic centre is given by 
\bea
\Phi^{tot}_{\gamma}(0) &\simeq& 1.8 \cdot 10^{-14} 
\;\left(\frac{2\,v\sigma_{2 \gamma} + v\sigma_{Z \gamma}}
{10^{-29}\ {\rm cm}^3 {\rm s}^{-1}}\right) 
\left(\frac{1\ {\rm TeV}}{M_\chi}\right)^2\;{\rm cm}^{-2}{\rm s}^{-1}{\rm sr}^{-1} \nonumber \\
&\simeq& 4\cdot 10^{-13}
\left({1\,{\rm TeV}\over M_\chi}\right)^2{\rm cm}^{-2}{\rm s}^{-1}{\rm sr}^{-1}.\label{eq:flx}
\eea

Although this flux is much higher than previous estimates have indicated,
it is still far from what is needed to explain the structure claimed
in \cite{strausz}.

The new generation of large area air Cherenkov telescopes may have a
sensitivity that comes close to the flux predicted in Eq.\,~(\ref{eq:flx}).
Moreover, exploiting their very small angular acceptance ($\sim 10^{-3} {\rm sr}$),
they will be able to investigate on the possible large central galactic density 
enhancement as predicted by the Navarro et al. profile, which  could dramatically increase the 
supersymmetric dark matter discovery potential of  these telescopes.

\section{Conclusions}

The results for the $Z\gamma$ process show many similarities with
the corresponding $2\gamma$ results. For heavy neutralinos, the two
lines would not be resolved so the 
line strengths would add. For neutralino masses below a few hundred
GeV, two distinct gamma lines could be detected permitting a ``spectroscopy''
which could give valuable  information on the nature (mass and composition)
of the dark matter particles. The analytical formulas given here should
be of great help to extract this  information, if these gamma
lines are detected.

Future gamma ray detectors may be sensitive enough to discover supersymmetric
dark matter for favourable values of supersymmetric and halo parameters, 
although the uncertainty of the latter may make exclusion of dark
matter models from the non-observation of a signal difficult.

\section{Acknowledgements}
We thank J. Edsj\"o and P. Gondolo for 
collaboration on the numerical supersymmetry calculations, and C. Frenk for
useful information on halo models.
The work of L.B. was supported by the Swedish Natural Science 
Research Council (NFR).

\begin{appendix}

\renewcommand{\thesection}{Appendix \Alph{section}}
\setcounter{equation}{0}
\renewcommand{\theequation}{\Alph{section}.\arabic{equation}}

\section{ }
In this Appendix, we define the functions needed to give the
expression for the cross section.
We start by defining the auxiliary functions
\beq
slog \left( r_1, r_2, r_3 ; x\right)\equiv 
\log\left[\left|-r_1 x^2 +\left(r_1+r_2-r_3\right) x +r_3\right|\right]
\eeq
and
\bea
\lefteqn{flog \left( r_1, r_2, r_3, r_4, r_5 ; x\right)\equiv 
\frac{1}{x -\left[\left(r_3+r_4\right)/2 -r_2-r_5\right]/
\left(r_1 +r_5\right)}\; \cdot } \nonumber \\
&&\cdot \left\{ \log\left[\left|-r_1 x^2 +\left(r_1-r_2+r_3\right) x 
+ r_2\right|\right] \right.\nonumber \\ 
&&\left.\;\;\;\;- \log\left[\left|r_5 x^2 + 1/2 \left(r_3-r_4\right) x 
+ 1/2 \left(r_3+r_4\right) - r_5\right|\right] \right\}.
\eea

These enter the following integrals, which cannot be performed analytically
in terms of elementary functions. We have chosen the integral expressions
rather than rewriting everything in terms of dilogarithms, because the
expressions are more compact and possible to integrate directly 
numerically.

\beq
I_1^{\;4}\left( r_1, r_2, r_3, r_4 \right) = \int_{0}^{1} \frac{d\,x}{x} 
\; \left[slog \left(-4\,r_1, r_2, r_3 ; x\right) 
- slog \left(-4\,r_4, r_2, r_3 ; x\right) \right] 
\eeq
\beq
I_1^{\;3}\left( r_1, r_2, r_3 \right) = 
I_1^{\;4}\left( r_1, r_2, r_2, r_3 \right) 
\eeq
\beq
I_1^{\;2}\left( r_1, r_2 \right) = I_1^{\;4}\left( r_1, r_2, r_2, 0 
\right) 
\equiv I_1\left( r_1, r_2 \right) 
\eeq

\beq
I_2^{\;4}\left( r_1, r_2, r_3, r_4 \right) = \int_{0}^{1} \frac{d\,x}{x} 
\; \left[slog \left(r_1-2\,r_4, r_2, r_3 ; x\right) 
- slog \left(-r_1, r_2, r_3 ; x\right) \right] 
\eeq
\beq
I_2^{\;3}\left( r_1, r_2, r_3 \right) 
= I_2^{\;4}\left( r_1, r_2, 1, r_3 \right) 
\eeq
\beq
I_2^{\;2}\left( r_1, r_2 \right) = I_2^{\;4}\left( r_1, r_2, 1, 0 \right) 
\equiv I_2\left( r_1, r_2 \right) 
\eeq

\beq
I_3^{\;4}\left( r_1, r_2, r_3, r_4 \right) = \int_{0}^{1} \frac{d\,x}{x} 
\; \left[slog \left(r_1-2\,r_4, r_3, r_2 ; x\right) 
- slog \left(-r_1, r_3, r_2 ; x\right) \right] 
\eeq
\beq
I_3^{\;3}\left( r_1, r_2, r_3 \right) 
= I_3^{\;4}\left( r_1, r_2, 1, r_3 \right) 
\eeq
\beq
I_3^{\;2}\left( r_1, r_2 \right) = I_3^{\;4}\left( r_1, r_2, 1, 0 \right) 
\equiv I_3\left( r_1, r_2 \right).
\eeq

$I_1\left( r_1, r_2 \right)$, $I_2\left( r_1, r_2 \right)$ and 
$I_3\left( r_1, r_2 \right)$ are the functions 
defined in Appendix A in Ref.\cite{lp}.
The notation here and in the following has been chosen such that 
when $M_Z\to 0$, the integrals have a simple relation to the ones in
\cite{lp}. For example, the $\tilde{I}^5$ function below also reduces
to $I_2$ or $I_3$ in the appropriate limits.

\bea
\tilde{I}^{\;5}\left( r_1, r_2, r_3, r_4, r_5 \right) &=& 
\int_{0}^{1}\, d\,x\,  \left[
\;flog \left(r_1-2\,r_5, r_2, r_3, r_4, r_5 ; x\right) \nonumber \right.\\
&&\;\;\;\;\;\;\;\;\;\;\;\;\left. \; 
- flog \left(-r_1, r_2, r_4, r_3, r_5 ; x\right) \;\right] 
\eea
\beq
\tilde{I}_{2;1}^{\;4}\left( r_1, r_2, r_3, r_4 \right) = 
\tilde{I}^{\;5}\left( r_1, r_2, r_3, 1, r_4 \right) 
\eeq
\beq
\tilde{I}_{2;2}^{\;4}\left( r_1, r_2, r_3, r_4 \right) = 
\tilde{I}^{\;5}\left( r_1, r_2, 1, r_3, r_4 \right) 
\eeq
\beq
\tilde{I}_{2}^{\;3}\left( r_1, r_2, r_3 \right) = 
\tilde{I}^{\;5}\left( r_1, r_2, 1, 1, r_3 \right) 
\eeq
\beq
\tilde{I}_{3}^{\;3}\left( r_1, r_2, r_3 \right) = 
\tilde{I}^{\;5}\left( r_1, 1, r_2, r_2, r_3 \right) 
\eeq

For the pieces that can be integrated analytically, it is convenient to
introduce the function

\begin{eqnarray}
K\left(r,\Delta \right) = \left\{
\begin{array}{ll} 
\frac{\sqrt{\Delta}}{2} \; \ln{\left(\left|\frac{\mbox{\normalsize 1}+(\mbox{\normalsize r}/\sqrt{\Delta})}
{\mbox{\normalsize 1}-(\mbox{\normalsize r}/\sqrt{\Delta})}\right|\right)} & \;\mbox{\normalsize if}\;
\; \Delta \geq 0 \\[4ex]
\sqrt{-\Delta} \;\arctan{\left(\frac{\mbox{\normalsize r}}{\sqrt{-\Delta}}\right)} 
& \;\mbox{\normalsize if}\;\; \Delta \leq 0 
\end{array}
\right.
\end{eqnarray}

\bea
I_{4}^{\;1}\left(a,b,c\right)& = &\frac{1}{a-c/4}\left[ K(1+a-b-c/2, 
\widetilde{\Delta}_1) - K(1-a-b+c/2, \widetilde{\Delta}_1 )\right. 
\nonumber \\
&& - K(1+a-b, \Delta_2 ) + K(1-a-b, \Delta_2) \nonumber \\ 
&& \left.+ K(c,c(c-4\,b)) + (1-b+c/4)\log{b}\, \right]
\eea
\bea
I_{4}^{\;2}\left(a,b,d,c\right)& = &\frac{1}{2\;(a-c/4)}\left[ 
K(1+a-b-c/2, 
\widetilde{\Delta}_1) + K(d+a-b-c/2, \widehat{\Delta}_1 )\right. 
\nonumber \\
&& - K(1-a-b+c/2, \widetilde{\Delta}_1) - K(d-a-b+c/2, \widehat{\Delta}_1 
) 
  \nonumber \\ 
&& - K(1+a-b, \Delta_2 ) - K(d+a-b, \widehat{\Delta}_2) + K(1-a-b, 
\Delta_2) 
\nonumber \\
&&+ K(d-a-b, \widehat{\Delta}_2) + K(c+1-d, \Delta_3) + K(c-1+d, \Delta_3) 
\nonumber \\
&&\left. + (1/2+d/2-b-c/4)\;(2\,\log{b}-\log{d}\,)\, \right]
\eea

 where we have defined
\bea 
\lefteqn{\widetilde{\Delta}_1\,=\,\left(a-c/2+b-1\right)^2 + 4\,(a-c/2)} 
\nonumber \\ 
\lefteqn{\widehat{\Delta}_1\,=\,\left(a-c/2+b-d\right)^2 + 4\,(a-c/2)\,d} 
\nonumber \\
\lefteqn{\Delta_2\,=\,\left(b-a-1\right)^2 - 4\,a} \nonumber \\
\lefteqn{\widehat{\Delta}_2\,=\,\left(b-a-d\right)^2 - 4\,a\,d} \nonumber \\
\lefteqn{\Delta_3\,=\,\left(1+d-c\right)^2 - 4\,d}
\eea 
Finally, for the imaginary parts we need the following functions:
\beq
J_{1}\left(a,b\right)=\log{\left(\frac{1+\sqrt{1-b/a}}{1-\sqrt{1-b/a}}
\right)}
\eeq
\beq
J_{2}\left(b,c\right)=\log{\left(\frac{1+\sqrt{1-4\,b/c}}{1-\sqrt{1-4\,b/c}}
\right)}
\eeq
\beq
J_{3}\left(a,b,c\right)=\log{\left( \frac{1-b+c/4+(a-c/4)\,\sqrt{1-4\,b/c}}
{1-b+c/4-(a-c/4)\,\sqrt{1-4\,b/c}}\right)}
\eeq

\end{appendix}

\pagebreak
\begin{figure}
 \centering
 \mbox{\subfigure{\epsfig{file=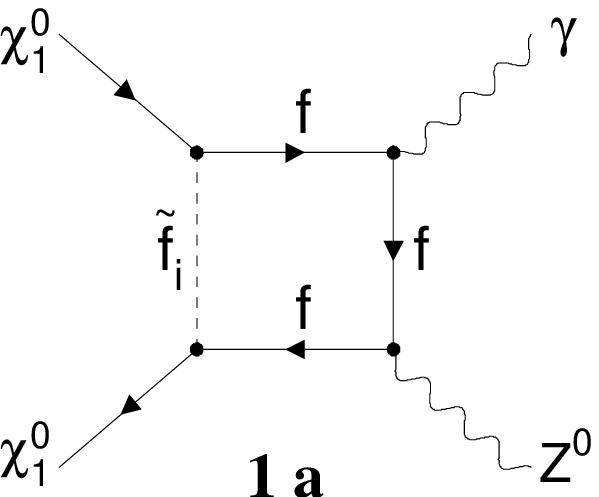,width=3cm}}\quad
       \subfigure{\epsfig{file=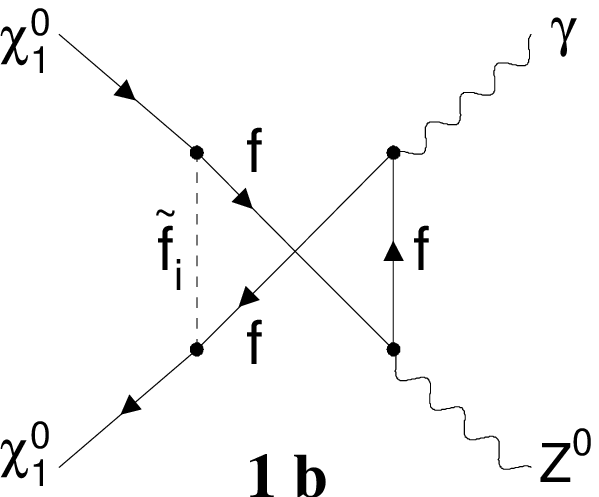,width=3cm}}\quad
       \subfigure{\epsfig{file=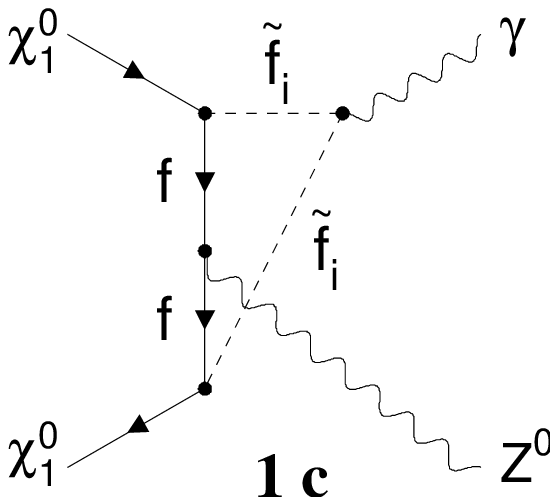,width=3cm}}\quad
       \subfigure{\epsfig{file=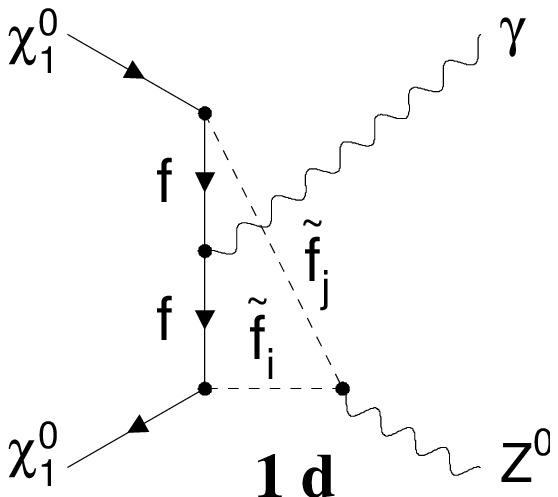,width=3cm}}}
 \mbox{\subfigure{\epsfig{file=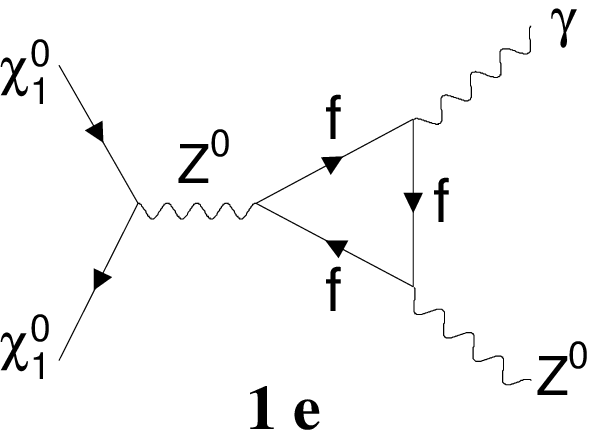,width=3cm}}\quad
       \subfigure{\epsfig{file=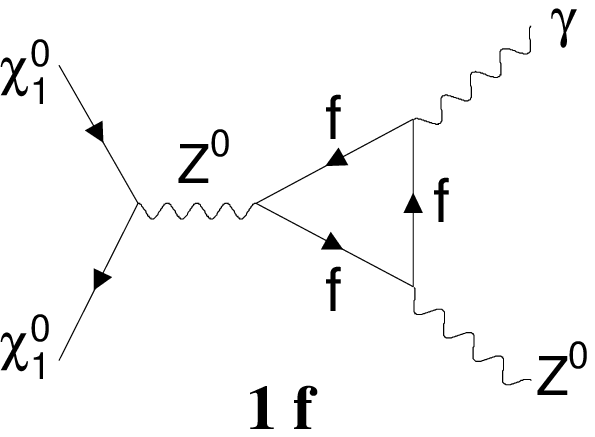,width=3cm}}\quad
       \subfigure{\epsfig{file=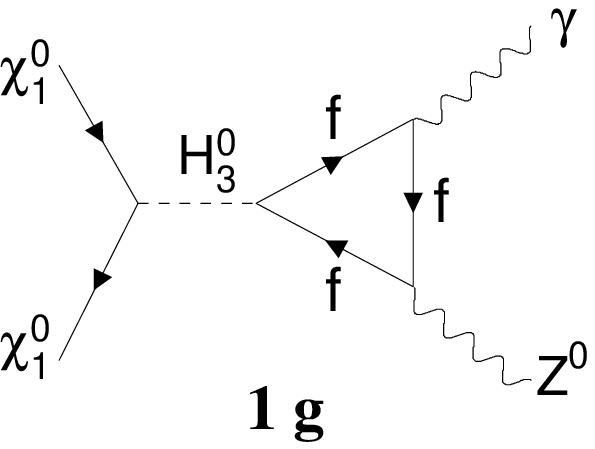,width=3cm}}\quad
       \subfigure{\epsfig{file=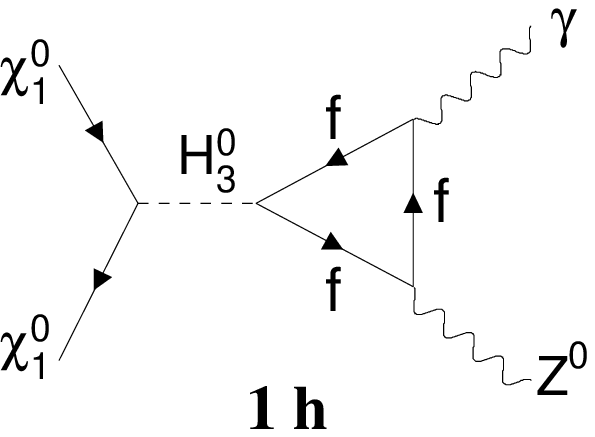,width=3cm}}}
\caption{Feynman diagrams included in the computation of $\tilde{\mathcal{A}}_{f\tilde{f}}$. Diagrams with exchanged initial 
states are not shown.}
\end{figure}

\begin{figure}
 \centering
 \mbox{\subfigure{\epsfig{file=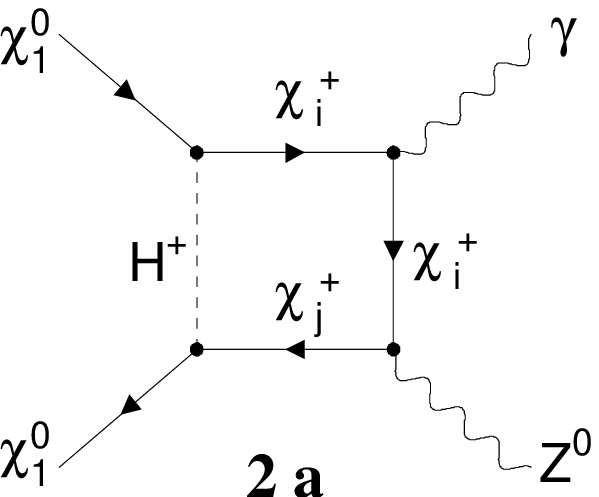,width=3cm}}\quad
       \subfigure{\epsfig{file=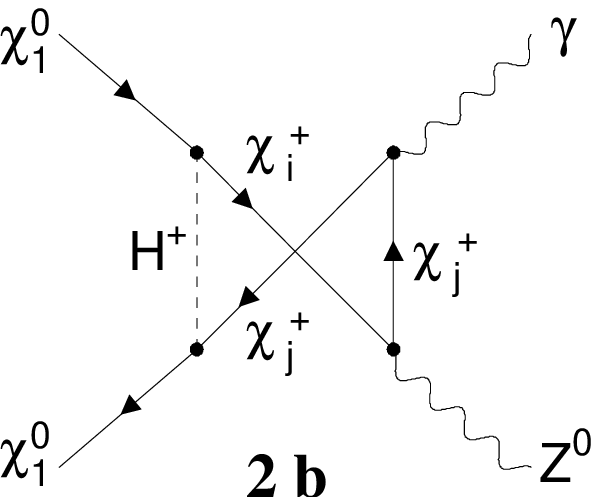,width=3cm}}\quad
       \subfigure{\epsfig{file=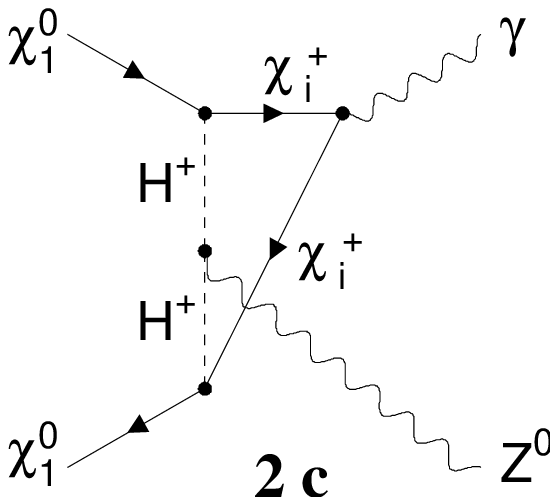,width=3cm}}\quad
       \subfigure{\epsfig{file=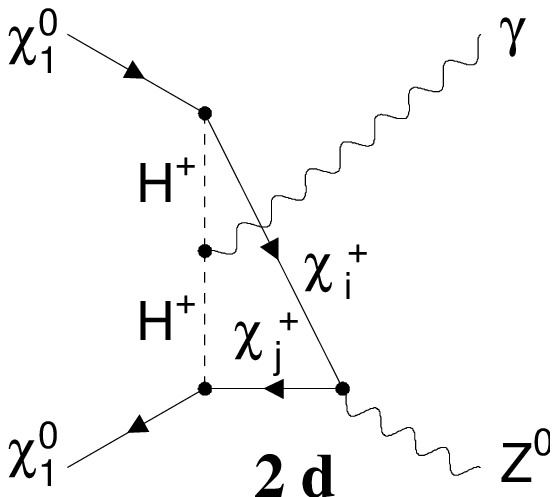,width=3cm}}}
 \mbox{\subfigure{\epsfig{file=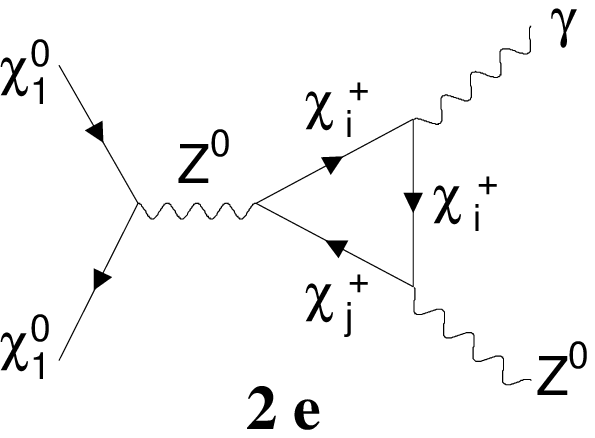,width=3cm}}\quad
       \subfigure{\epsfig{file=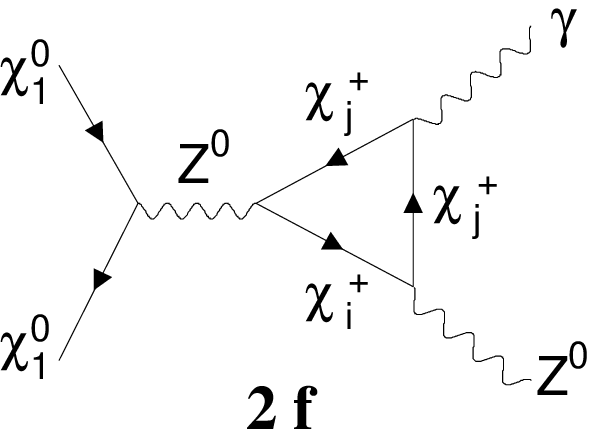,width=3cm}}\quad
       \subfigure{\epsfig{file=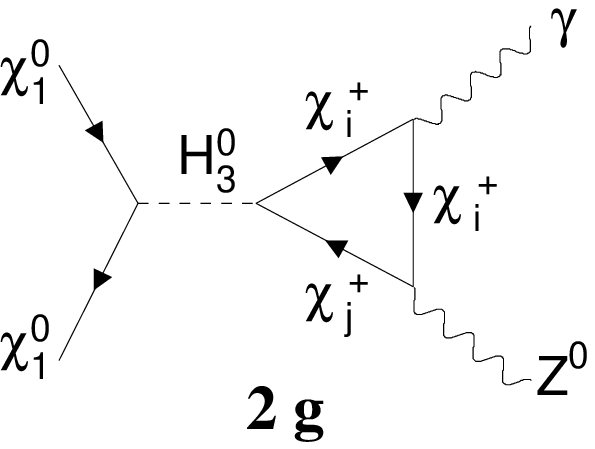,width=3cm}}\quad
       \subfigure{\epsfig{file=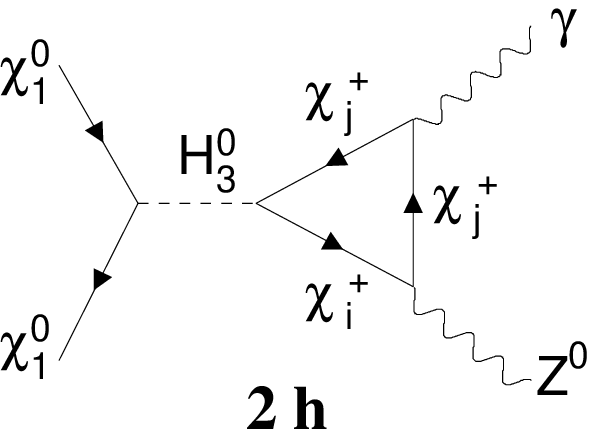,width=3cm}}}
\caption{Feynman diagrams included in the computation of $\tilde{\mathcal{A}}_{H^+}$ . Diagrams with exchanged initial 
states are not shown.}
\end{figure}

\begin{figure}
 \centering
 \mbox{\subfigure{\epsfig{file=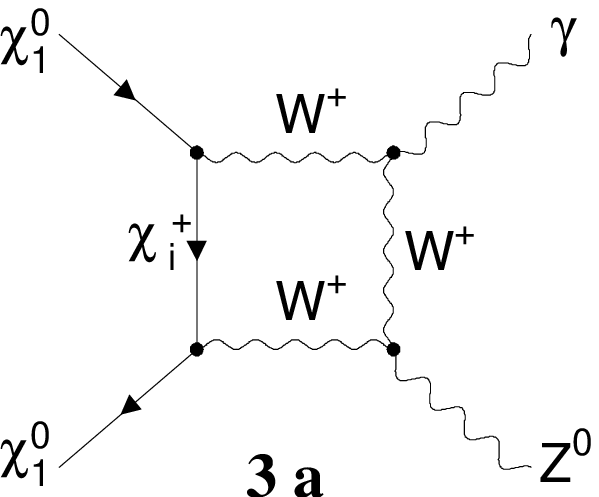,width=3cm}}\quad
       \subfigure{\epsfig{file=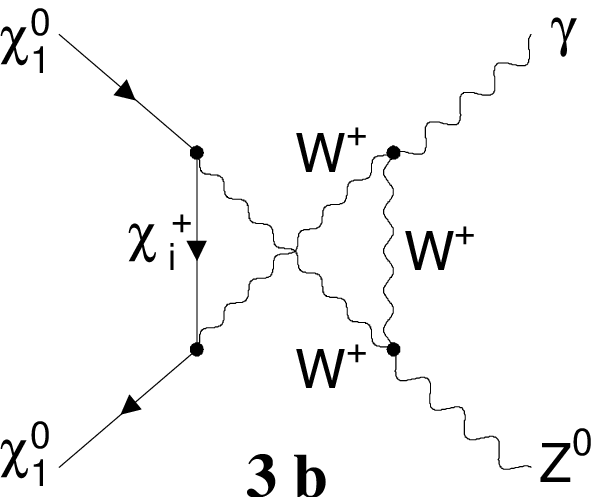,width=3cm}}\quad
       \subfigure{\epsfig{file=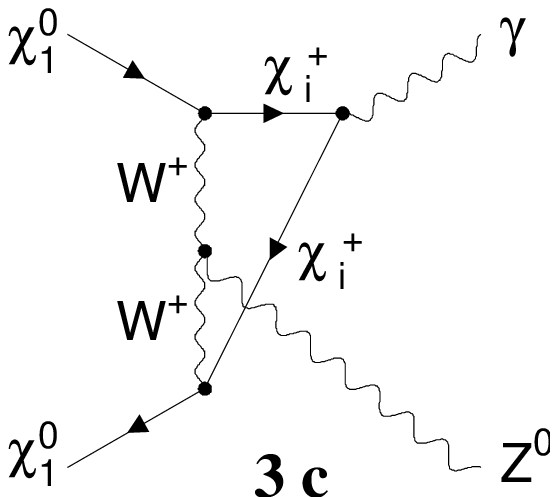,width=3cm}}}
 \mbox{\subfigure{\epsfig{file=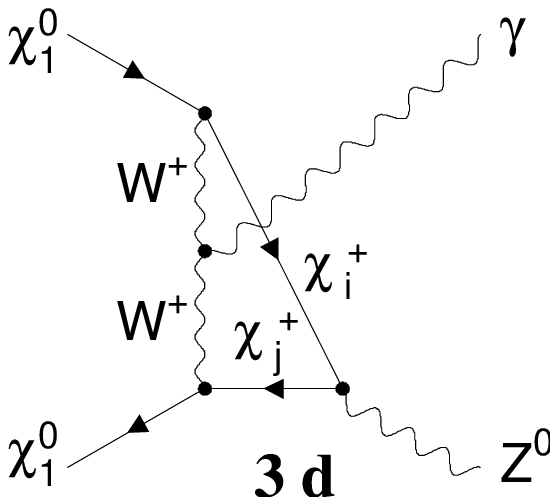,width=3cm}}\quad
       \subfigure{\epsfig{file=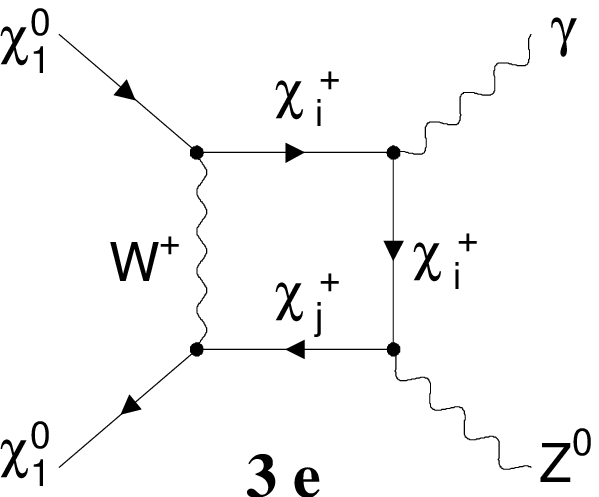,width=3cm}}\quad
       \subfigure{\epsfig{file=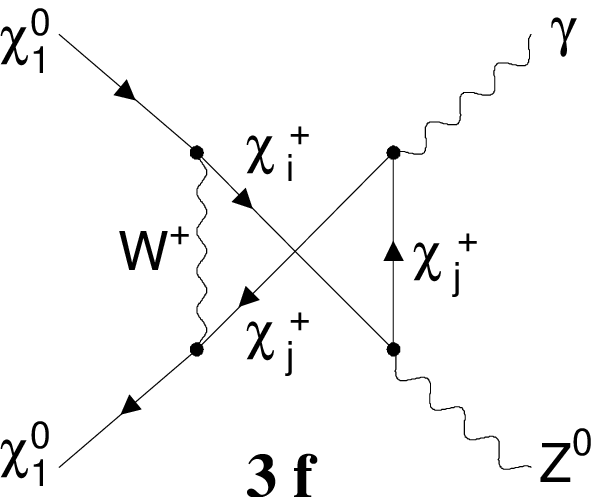,width=3cm}}}
\caption{Feynman diagrams included in the computation of $\tilde{\mathcal{A}}_
{W}$ . Diagrams with exchanged initial  
states are not shown.}\label{fig:fig1}
\end{figure}

\begin{figure}
 \centering
 \mbox{\subfigure{\epsfig{file=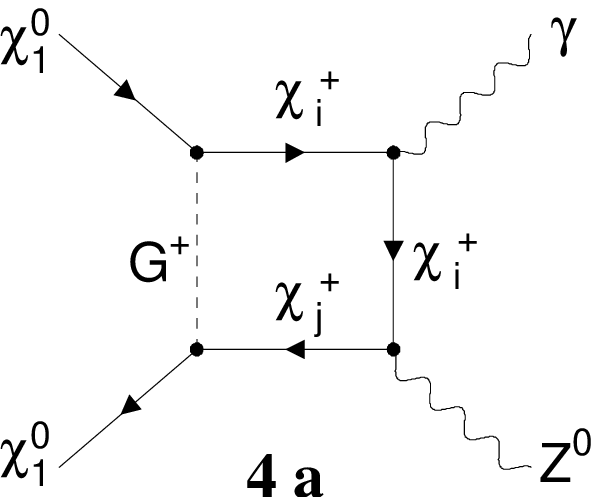,width=3cm}}\quad
       \subfigure{\epsfig{file=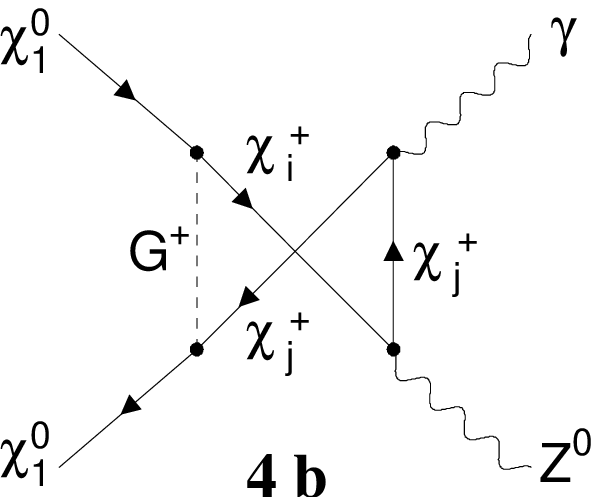,width=3cm}}\quad
       \subfigure{\epsfig{file=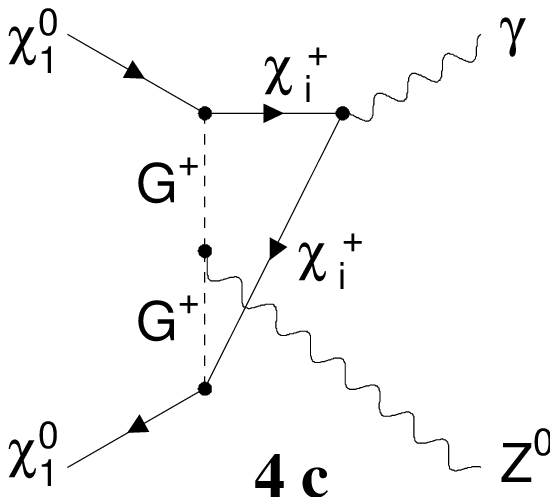,width=3cm}}}
 \mbox{\subfigure{\epsfig{file=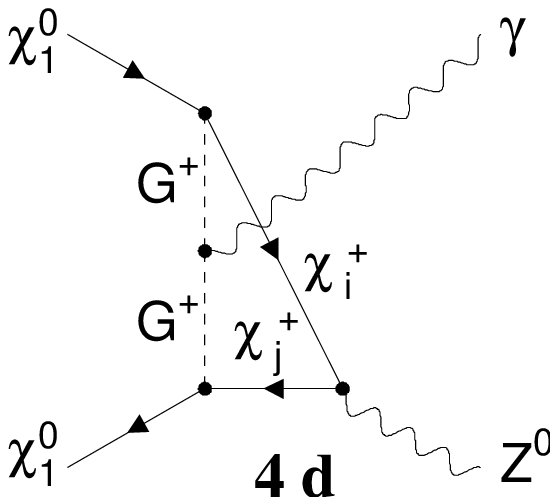,width=3cm}}\quad
       \subfigure{\epsfig{file=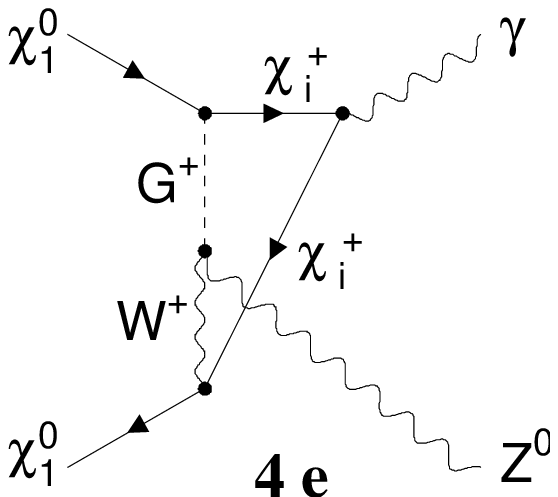,width=3cm}}\quad
       \subfigure{\epsfig{file=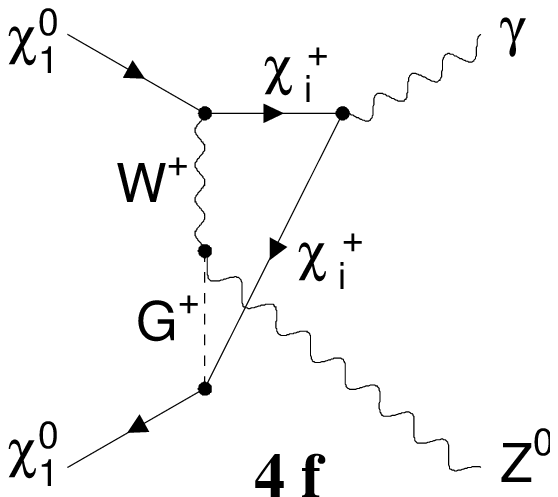,width=3cm}}}
\caption{Feynman diagrams included in the computation of $\tilde{\mathcal{A}}_{G}$. Diagrams with exchanged initial 
states are not shown.}
\end{figure}

\begin{figure}
 \centering
 \mbox{\subfigure{\epsfig{file=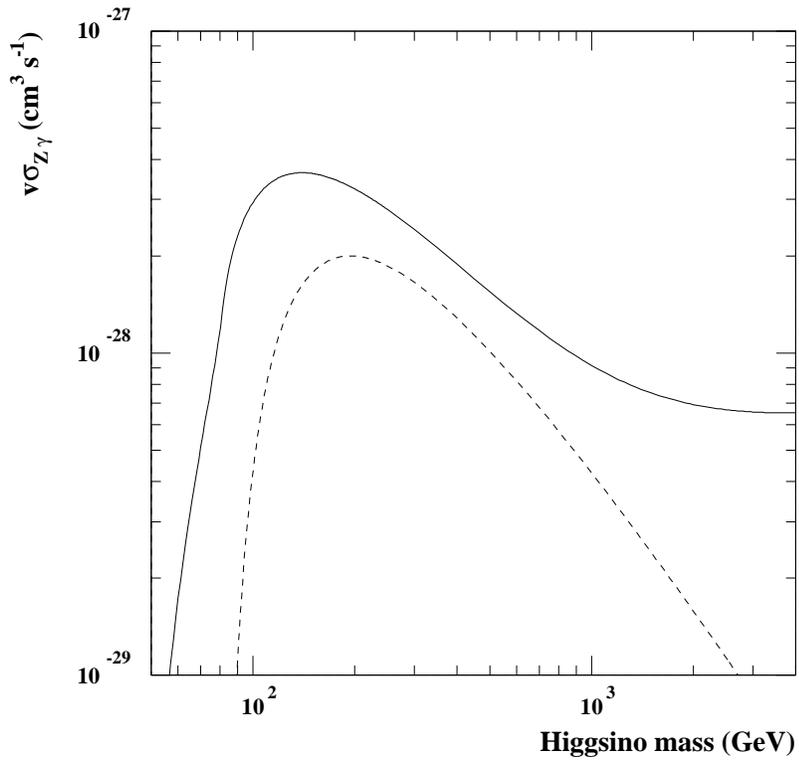,width=12cm}}}
\caption{Annihilation rate of pure higgsinos into a photon and a Z$^0$ boson obtained in this work (solid line). Also shown is the  unitary bound 
coming from keeping the imaginary part only, which agrees with the result
of Ref.~[4] (dashed line).}
\end{figure}

\begin{figure}
 \centering
 \mbox{\subfigure{\epsfig{file=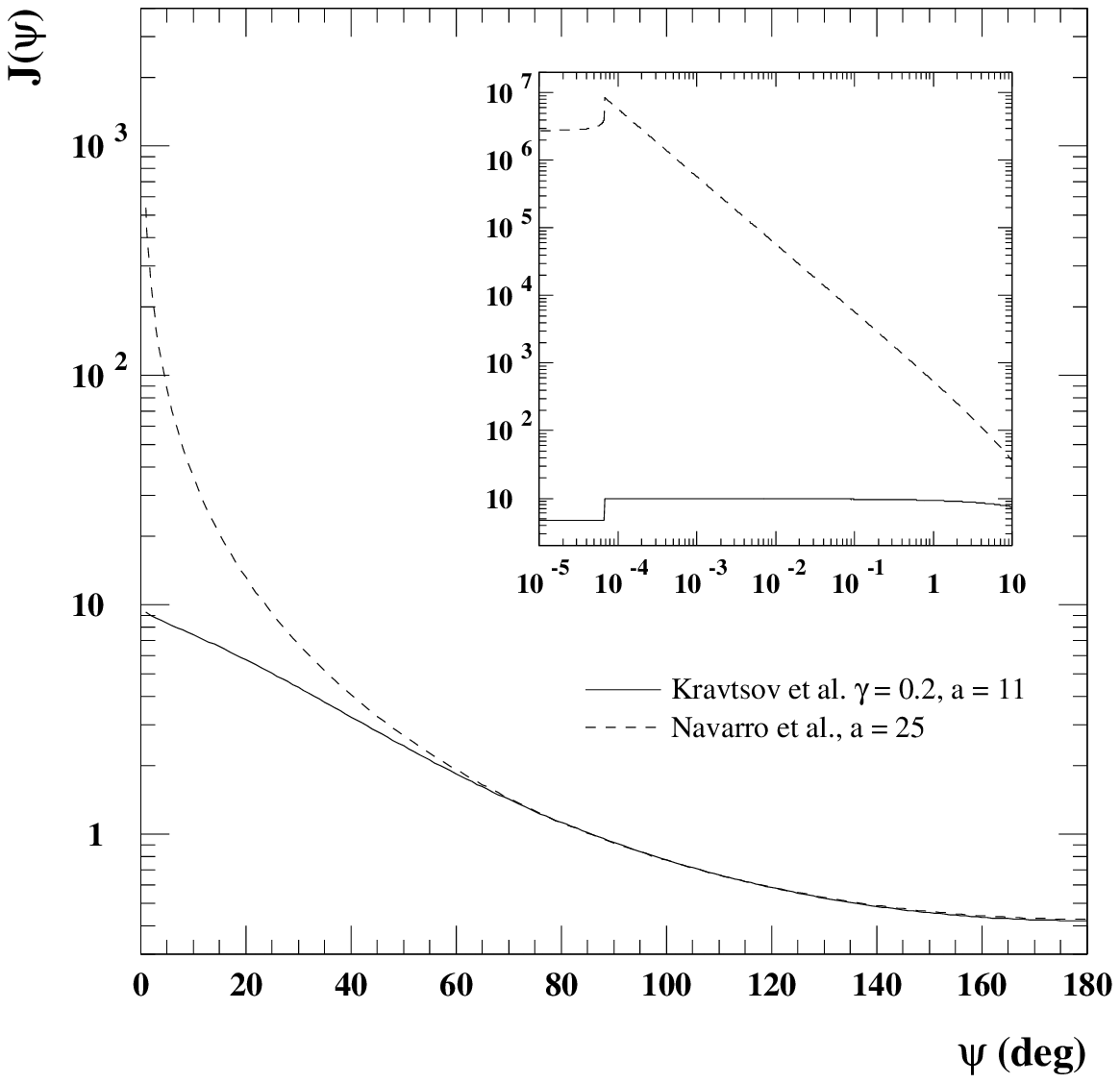,width=12cm}}}
\caption{Function $J(\psi)$ as defined in Eq.~(18). In the detail the discontinuity in the two curves is due to the portion of halo which is hidden by the black hole at the galactic centre.}
\end{figure}


\begin{thebibliography}{99}

\bibitem{jkg} G. Jungman, M. Kamionkowski and K. Griest, Phys. Rep.
{\bf 267} (1996) 195.

\bibitem{lp} L.~Bergstr{\"o}m and P.~Ullio, hep-ph/9706232.

\bibitem{strausz} S.C. Strausz, Phys. Rev. {\bf D55} (1997) 4566. 

\bibitem{BK}
L.~Bergstr{\"o}m and J.~Kaplan, Astropart.\ Phys. {\bf 2} (1994) 261.

\bibitem{STS} M. Srednicki, S. Theisen and J. Silk, Phys. Rev. Lett. 
{\bf 56} (1986) 263.

\bibitem{BS} L. Bergstr\"om and H. Snellman, Phys. Rev.
{\bf D37} (1988) 3737.

\bibitem{kuhn} 
J.H.~K\"{u}hn, J.~Kaplan and O.~Safiani, Nucl.\ Phys.\ {\bf B157} 
(1979) 125. 
\bibitem{fujikawa}
K.~Fujikawa, Phys.\ Rev. {\bf D7} (1973) 393.

\bibitem{joakim} 
J.~Edsj{\"o}, Aspects of Neutrino Detection of Neutralino Dark Matter 
(Uppsala University thesis, Uppsala, 1997), hep-ph/9704384. 

\bibitem{zuber} 
C.~Itzykson and J.B.~Zuber, Quantum Field Theory (McGraw-Hill, Singapore, 
1980). 

\bibitem{joakimpaolo} J. Edsj\"o and P. Gondolo, hep-ph/9704361 (1997).

\bibitem{binney} W. Dehnen and J. Binney, astro-ph/9612059 (1997).

\bibitem{kravtsov} A. V. Kravtsov et al., astro-ph/9708176 (1997).

\bibitem{navarro}J.F. Navarro, C.S. Frenk and S.D.M. White, Astrophys. J.
{\bf 462} (1996) 563.

\bibitem{bere} V.S.~Berezinsky, A.V.~Gurevich and K.P.~Zybin, 
Phys. Lett. {\bf B294} (1992) 221.
\bibitem{flores} R.A. Flores and J.R. Primack, Astrophys. J. {\bf 427} (1994)
L1.

\bibitem{bs} A. Burkert and J. Silk, astro-ph/9707343 (1997).

\bibitem{frenk_priv} C. Frenk, private communication.

\bibitem{turner} M.S. Turner, Phys.\ Rev. {\bf D34} (1986) 1921.
\end{thebibliography}
\end{document}